# Direct fabrication of three-dimensional buried conductive channels in single crystal diamond with ion microbeam induced graphitization

P. Olivero[1,2,3]*, G. Amato[4], F. Bellotti[4], S. Borini[4], O. Budnyk[1,3], E. Colombo[1,3], M. Jakšić[2], A. Lo Giudice[1,3], C. Manfredotti[1,3], Ž. Pastuović[2], F. Picollo[3], N. Skukan[2], M. Vannoni[5], E. Vittone[1,3]

[1] Centre of Excellence "Nanostructured Interfaces and Surfaces", University of Torino, Via P. Giuria 7, 10125 Torino, Italy

[2] Laboratory for Ion Beam Interactions, Ruđer Bošković Institute, Bijenička 54, HR-10000 Zagreb, Croatia

[3] Experimental Physics Department, University of Torino and INFN sez. Torino, Via P. Giuria 1, 10125 Torino, Italy

[4] Quantum Research Laboratory, Istituto Nazionale di Ricerca Metrologica, Strada delle Cacce 91, 10135 Torino, Italy

[5] Istituto Nazionale di Ottica Applicata, CNR, Largo E. Fermi 6, Arcetri, 50125 Firenze, Italy

* corresponding author

**Abstract**

We report on a novel method for the fabrication of three-dimensional buried graphitic micropaths in single crystal diamond with the employment of focused MeV ions. The use of implantation masks with graded thickness at the sub-micrometer scale allows the formation of conductive channels which are embedded in the insulating matrix at controllable depths. In particular, the modulation of the channels depth at their endpoints allows the surface contacting of the channel terminations with no need of further fabrication stages.

In the present work we describe the sample masking, which includes the deposition of semi-spherical gold contacts on the sample surface, followed by MeV ion implantation. Because of the significant difference between the densities of pristine and amorphous or graphitized diamond, the formation of buried channels has a relevant mechanical effect on the diamond structure, causing localized surface swelling, which has been measured both with interferometric profilometry and atomic force microscopy. The electrical properties of the buried channels are then measured with a two point probe station: clear evidence is given that only the terminal points of the channels are electrically connected with the surface, while the rest of the channels extends below the surface. IV measurements are employed also to qualitatively investigate the electrical properties of the channels as a function of implantation fluence and annealing.

**Introduction**

The process of graphitization induced in diamond by ion implantation was first investigated by Vavilov et al. in the 70s [1]. This pioneering work triggered a series of studies on the effects of ion induced damage on the electrical transport properties of diamond. Hauser et al. demonstrated that the electrical properties of ion-implanted diamond layers were similar to those of amorphous carbon produced by sputtering graphite [2, 3]. The hopping conduction in diamond implanted with carbon ions at different energies and fluences was investigated by Prins [4, 5], who interpreted the onset for this process with mechanisms of vacancy-interstitial interaction, focusing on the different mobility of vacancy and interstitial defects in the crystal. In a series of Ar and C implantation experiments carried at different temperatures, Sato et al. demonstrated that target temperature during implantation has a strong influence on the ion damage processes that determine the increase in conductivity, as confirmed by Raman characterization [6, 7]. The effect of C and Xe ion induced graphitization in polycrystalline samples grown by chemical vapor deposition was studied by S. Prawer et al., demonstrating that the fluence dependence of the electrical conductivity of the implanted area is similar to what measured in single crystal diamond [8]. Selective Co ion implantation on self-supporting diamond films was employed by B. Miller et al. to pattern conductive areas on which subsequent redox electron transfer and metal deposition were demonstrated [9]. S. Prawer et al. performed IV measurements *in-situ* during C and Xe implantation at different temperatures, showing complex non-monotonic dependencies of the electrical conductivity from the ion fluence. These trends confirm that the critical fluence at which a sharp decrease in conductivity (due to the formation of a continuous conducting pathway) is observed strongly depends from the implantation temperature [10, 11]. B implantation studies carried by F. Fontaine et al. on polycrystalline diamond [12] confirmed the basic interpretation of the process, while introducing two different critical fluences. In their interpretation, when a "percolative threshold" fluence is reached, a continuous conductive path of $sp^3$-like defects is established in the implanted material and variable range hopping conduction appears, while at fluences above a slightly higher "amorphization threshold" a network of $sp^2$-bonded defects is formed, which leads to the permanent graphitization of the implanted areas upon thermal annealing. F. Prins carried further theoretical calculations on the data reported in [11] to interpret the onset for variable-range-hopping in a model that does not include the formation of displacement spikes in diamond crystal [13]. An extensive IV characterization in temperature of diamond implanted with Xe ions at low temperature was carried by A. Reznik et al. [14, 15], allowing the extraction of a number of characteristic energies for hopping sites from the temperature dependence of the resistivity [16].

While the interpretation of the onset of hopping-related conduction mechanisms at low damage densities has been debated in detail in the above mentioned works, the formation of ohmic conductive paths in diamond at high damage densities has a more straightforward interpretation based on the formation of a stable graphite-like $sp^2$ network, and also found several interesting applications. S. Prawer et al. demonstrated that electrically heated resistive paths created with carbon implantation can act as infrared radiation emitters, in which IR emission is confined to the conductive damaged areas [17]. A. V. Karabutov et al. employed nitrogen implantation followed by thermal annealing to provide surface electrical conductivity to diamond microtips and improve their performance as field emitters [18, 19]. Moreover, the possibility of creating effective ohmic contacts on doped or intrinsic diamond by high fluence implantation has been explored in several works [20, 21, 22, 23, 24].

Interestingly, while this research topic can be considered mature in terms of fundamental studies and technological applications, the possibility of fabricating buried graphitic channels, i.e. conductive paths extending below the diamond surface, has not been fully explored, with the exception of the works reported by R. Walker et al. [25, 26] and by E. Trajkov et al. [27]. In the former works, buried conductive layers were created with 2 MeV B implantation; the implantation fluence was kept below the graphitization threshold in order to achieve doping-related p-type conduction, although the buried channels were contacted with the sample surface with laser-induced graphitization. In the latter work, 2 MeV He ions were implanted in order to measure defect-related conductivity in the sub-graphitization range.

It is worth stressing that, with the exception of the above mentioned works, the implantation of relatively heavy ions (C, Ar, Xe, Co, B, N) at energies of few hundreds of keV has been employed so far, thus leading to the formation of conductive paths and contacts at the sample surface. In the present work we report for the first time, to our knowledge, on the formation and electrical characterization of buried graphitic paths formed in diamond with MeV ion implantation.

**Ion implantation in diamond**

The process of damage induced by energetic ions in matter occurs mainly at the end of ion range, where the cross section for nuclear collisions is strongly enhanced, after the ion energy is progressively reduced by electronic collisions occurring in the initial stages of the ion path [28]. The permanent conversion of ion-implanted diamond to a graphite-like phase upon thermal annealing occurs when a critical damage density (usually referred as "graphitization threshold") is reached. Such threshold value has been estimated as $1 \cdot 10^{22}$ vacancies $cm^{-3}$ for shallow implantations by C. Uzan-Saguy et al. [29] and as $9 \cdot 10^{22}$ vacancies $cm^{-3}$ for deep implantations by P. Olivero et al. [30]. The discrepancy between the two values has been attributed in [30] to the higher internal pressure for deep implantations which could effectively increase the graphitization threshold, as already suggested in previous works [31, 32].

In the range of keV ion implantation conditions applied in the above mentioned works, when the critical fluence is reached for the onset of the graphitization process, the damaged layer extends from the end of range of ions to the surface of the sample. On the other hand, MeV implantation reported in the above mentioned works do not reach the graphitization threshold. This is shown in the TRIM [33] Monte Carlo simulation in Fig. 1 for the implantation conditions reported in [6, 11, 12, 25, 27]. The curves were calculated by setting a value of 50 eV for the atom displacement energy in the diamond lattice [34, 35]. In Fig. 1 the damage profile of 6 MeV carbon ions in diamond is also shown; for an implantation fluence of $1.5 \cdot 10^{16}$ $cm^{-2}$, the higher value of the graphitization threshold in the bulk material determines the formation of a narrow damaged layer at a depth of ~2.75 μm below the surface, which can permanently convert to a graphite-like phase upon thermal annealing. On this basic concept the "diamond lift-off" process was developed by N. P. Parikh et al. [36].

In order to connect the endpoints of the channels to the sample surface, a three-dimensional masking technique was developed to modulate the penetration depth of the ions from their range in the unmasked material up to the sample surface with increasing thickness of stopping material. The basic concept is shown schematically in Fig. 2. The ion implantation is performed by scanning an ion beam along a linear path, which terminates at a semi-spherical mask characterized by a non-uniform thickness profile (Fig. 2a). As the beam scan progresses towards the center of the mask, incident ions cross an increasing thickness of masking material, thus progressively reducing their range in the diamond layer. Therefore, after the removal of the mask, the heavily damaged layer is connected with the sample surface (Fig. 2b) at the position where the mask had a thickness equal to the ion range in the masking material. This simple approach has the advantage of being completely integrated with the implantation process, provided that a suitable mask is positioned on the sample surface. Moreover, a fine control over the mask thickness is not strictly required: as long as the mask is characterized by a sloping profile along its edges and it is thick enough in its central region, it is expected that at some point within its extension the damaged layer will be connected to the surface. If the material of which it is composed is conductive, the mask itself can be used as a superficial electrical contact point with the graphitic path.

**Experimental**

The sample under exam is a synthetic single crystal diamond produced by Sumitomo with high pressure high temperature (HPHT) method. The crystal is 3×3×1.5 mm in size and it is classified as type Ib, having a substitutional nitrogen concentration between 10 and 100 ppm, as indicated by the manufacturer. The sample is cut along the 100 crystal direction and it is optically polished on the two opposite large faces.

Before ion implantation, the sample was masked with semi-spherical contacts by means of a standard gold wire ball bonder commonly used for the contacting of microchip devices. Since the adhesion of gold contacts on a polished diamond surface is extremely poor, four gold contacts were deposited in a square shape after having evaporated four Cr-Au adhesion areas on the sample surface through a mask, labeled as A, B, C, D in Fig. 3. The Cr layer was evaporated while keeping the sample at a temperature of 300 ºC in order to improve its adhesion to the diamond surface. The Cr and Au adhesion layers are ≈10 nm and ≈60 nm thick, respectively, therefore they do not affect significantly the depth at which the damaged layer is formed in diamond; on the other hand, the gold semispherical maps have a thickness of ~50 μm in their central regions, which is more than enough to fully stop the incident ions.

After sample masking, 6 MeV $C^{3+}$ ions were implanted at room temperature at the microbeam line of the Laboratory for Ion Beam Interaction of the Ruđer Bošković Institute. The ion current was between 100 pA and 300 pA, and the ion were focused to a micrometer-size spot which was raster scanned to achieve a uniform fluence delivery across the implantation area. Three strip areas connecting the mask contacts were implanted at increasing fluences, as shown in Fig. 4a. The Rutherford Backscattering signal was employed to locate the golden contacts and adhesion layers on the sample surface, thus allowing the definition of the implanted areas with good spatial accuracy. The implantation fluences for channels 1, 2 and 3 were $6.3 \cdot 10^{15}$ cm$^{-2}$, $1.1 \cdot 10^{16}$ cm$^{-2}$ and $1.7 \cdot 10^{16}$ cm$^{-2}$, respectively, as evaluated from the implantation time and ion current on known areas.

After ion implantation, the sample was annealed at a temperature of 800 ºC for 1 hr in vacuum with the purpose of graphitizing the heavily damaged layers at the end of ion range, while recovering the pristine diamond structure in the superficial layers where a lower damage density was created. As reported in previous works [37], ion implantation in diamond is associated with surface swelling which is due to the significant difference in density between diamond ($\rho_D = 3.52$ g cm$^{-3}$), amorphous carbon and graphite ($\rho_G = 2.16$ g cm$^{-3}$). For deep implantations, the swelling is mainly determined by the volume expansion of the heavily damaged layer at the end of range of ions, which exerts a significant pressure on the surrounding and relatively intact diamond lattice. Both atomic force and interferometric profilometry were carried to measure accurately the surface swelling associated with the channel implantation. The electrical characterization of the conductive channels was performed before and after the thermal annealing process, as described in the following section.

**Morphological characterization**

The channels morphology was studied with White-Light Interferometric (WLI) profilometry and Atomic Force Microscopy (AFM), with the purpose of characterizing the damage-induced swelling of the irradiated areas.

The measurements of white-light interferometric profilometry were carried at the Optical Testing Laboratory of the "Istituto Nazionale di Ottica Applicata" (INOA-CNR), under controlled environmental conditions (clean-room class <10,000). A commercial system (NewView 5700 by Zygo Corporation) was employed: using the interference between the beam reflected by the diamond surface and a reference beam, a three dimensional profile of the surface was obtained by raster-scanning the sample with a nanometer accuracy. The advantage of using low-coherence interferometry is that the measurement is not affected by other external reflections (as for example, the back diamond surface) [38, 39]. Figs. 5a shows the profilometry map of channel #2 after thermal annealing; a pronounced swelling of 145 nm is localized at the implanted area and clearly visible. It is worth noting that the width of the swollen area, as measured from the cross-sectional profile reported in Fig. 5b, is similar to the channel width reported in Fig. 4b (i.e. ~16 μm). This indicates that the diamond lattice surrounding the buried channel reacts with surprising flexibility to the high internal pressures associated with the volume expansion of the damaged areas. Similar maps and profiles were collected for channels #1 and #3, yielding swelling thicknesses of 20 nm and 195 nm, respectively. As expected, the swelling thickness increases with increasing fluence.

AFM mapping was employed to investigate the swelling profile of the channels at their endpoints, where the buried channels emerge from the bulk and connect with the sample surface. A Park System XE100 AFM was employed in to measure the swelling profiles in contact, tapping and non-contact modes, yielding consistent results. Fig. 6a shows the AFM map collected in tapping mode of the endpoint of a different channel obtained with implantation conditions similar to those described above for channel #3, after the removal of the three-dimensional mask and before thermal annealing. As shown in the transverse profile reported in Fig. 6b, the swelling thickness away from the channel endpoint is 200 nm, while interestingly the swelling increases up to a value of 300 nm at the endpoint, where the channel is emerging at the surface (see Fig,. 6c). Such a significant increase is highlighted in the longitudinal profile reported in Fig. 6d, and can be qualitatively explained if it is considered that, as the channel approaches the sample surface, the mechanical forces arising from the surrounding diamond lattice that keep the damaged layer under compression are progressively reduced. A detailed quantitative analysis of the variation of the swelling thickness as a function of the implantation depth is beyond the scope of the present paper and will be carried in future works. Nonetheless, it is worth stressing that the measured increasing swelling at decreasing implantation depths strongly suggests that, because of the high internal pressures associated with the volume expansion of the damaged layers within a rigid crystal lattice, the process of ion-induced damage by deep implantation is significantly inhibited in comparison to shallow implantations.

**Electrical characterization**

The electrical conduction properties of the buried channels were measured with a 4145B Semiconductor Parameter Analyzer by Hewlett Packard; two tip probes were employed in voltage source mode. The sensing tips probed the electrical conductivity of the buried channels at the surface metal contacts A, B, C, D. Test measurements were also performed by positioning the probes at two points (E, F) directly above channel #3 on the unmasked diamond surface (see Fig. 4a). The results of the measurements are reported in Figs. 7a,b, showing respectively the IV characteristic of the channels in linear-linear and linear-logarithmic scales. The use of the two different scales allows the identification of a non-linear conductivity trend versus the bias voltage (Fig. 7a), and the appreciation of the extremely low values of the current measured at the test points (Fig. 7b).

The test measurements confirm very convincingly that the conductivity is to be attributed to buried channels that are completely disconnected from the sample surface, with the exception of their endpoints which are in electrical contact with the surface electrodes. The IV curves follow a super-linear trend which has two compatible explanations: charge injection mechanisms at the interface between the channels and the electrodes, and non-ohmic charge transport mechanisms in the channels. With regards to the latter point, it is worth stressing that before thermal annealing the channels consist of disordered networks of $sp^2$ and $sp^3$ chemical bonds which are still to be re-arranged in a continuous graphite-like structure. As expected, the channel implanted at the highest fluence (i.e. channel #3) displays the highest conductivity, while no conductivity measured above the noise level from channel #1.

Fig. 8 reports the results of IV characterization of the channels after thermal annealing. In Fig. 8a the rectification of the IV characteristics for channels #3 and #2 is clearly visible, indicating ohmic conduction in the heavily damaged channels. An increase in conductivity of $\sim 10^2$ and $\sim 10^3$ is observed for channels #3 and #2, respectively, while the conductivity of channel #1 is still not visible in the linear scale of Fig. 8a. In the linear-logarithmic plot of Fig. 8b the increase in conductivity of channel #1 can be appreciated. Such an increase is associated to an analogous (although less significant) increase measured from test points (E, F), which is attributed to surface contamination of the sample surface from the electrical contacts during the thermal annealing, as confirmed by similar conductivity values measured on the unmasked diamond surface at different points which are not located directly above the channels. The surface conductivity is not localized above the channels, and the conductivity of channels #3 and #2 is well above this level, therefore we attribute it to ohmic conduction in the buried channels. On the contrary, the increase in conductivity between points A and B cannot be unequivocally attributed to an increase in conductivity of channel #1 with respect to superficial parasite currents.

The above mentioned measurements were confirmed by removing the masking contacts, cleaning the sample surface in acetone and re-contacting with evaporated metal contacts.

The measurements are not reported here, and it is worth noticing that the sample could not be cleaned more accurately (i.e. with an acid chemical attack), in order to avoid the etching of the exposed graphite [30].

These trends can be qualitatively interpreted if the damage density profiles for the three implantations are considered in comparison to the threshold damage density reported in [30]. As shown in Fig. 9, at the highest implantation fluence (channel #3) the end-of-range damage peak is significantly higher than the graphitization threshold, therefore it is expected that a narrow graphite-like buried path is formed upon thermal annealing. The implantation fluence of channel #2 reaches the graphitization limit in a much narrower region, while for channel #1 the implantation fluence is well below the critical threshold. As reported in [30], the accurate determination of the graphitization threshold from the interpolation of the layer thickness on the calculated damage profile was difficult because of the high slope of such profile near the end of range. Therefore, the value of the critical damage density for deep implantations is affected by a significant experimental uncertainty. Nonetheless, the IV data reported here confirm such value is significantly higher than what reported for shallow implantations, and that assuming a factor of ~10 between the two values is not unreasonable.

It is worth noticing that the surprisingly low conductivity of channel #1 is confirmed by optical measurements in transmission carried on the annealed sample after the removal of the mask electrodes. As shown in Fig. 10, in the annealed sample, channel #1 becomes more transparent, in comparison to channel #2 that still retains its opacity. Such a phenomenon strongly indicates the partial recovery of the diamond structure in a channel implanted at a fluence below the critical threshold.

Estimating the resistivity of channels #1 and #2 finds a significant obstacle in the evaluation of their thickness, while the evaluation of the other geometrical parameters (length and width) is possible from optical measurements in transmission, as shown in Figs. 4, 10. It is possible to estimate the channel thicknesses from the interpolation of the calculated damage profiles reported in Fig. 9 with the graphitization threshold. Such an approach does not yield consistent results, which we attribute to the uncertainty on the value of the graphitization threshold. On the other hand, it is possible to use the swelling thicknesses resulting from interferometric profilometry as an estimation of the lower boundary for the thickness of the damaged layer. If such values are used together with the other geometrical parameters and the resistance values evaluated at V = 80 V, resistivities of 0.9 Ω cm and 1.1 Ω cm for channels #2 and #3 after thermal annealing are obtained, as summarized in Table 1. Such values are in satisfactory agreement, indicating that the significantly different resistance of the two channels after thermal annealing is to be primarily attributed to the different geometrical parameters, while the conducting medium in the channels are characterized by a very similar resistivity. It is worth stressing that such values represent the lowest estimation of the real resistivities, since the channel thicknesses were probably under-estimated by using the swelling thicknesses. Nonetheless, the values are well above the resistivity of graphite ($1.4 \cdot 10^{-3}$ Ω cm), suggesting that the conducting medium is not constituted of an ordered $sp^2$-bonded phase. For this reason, the optimization of the implantation process and a more detailed electrical characterization of the buried channels will be investigated in future works.

**Conclusions**

In the present work we present the first demonstration of monolithical fabrication and morphological/electrical characterization of buried graphitic channels in diamond with MeV ion implantation through three-dimensional masks followed by thermal annealing.

The morphological characterization of the channels shows that the swelling associated with deep ion implantation increases towards the endpoints of the channels, indicating that the mechanical forces arising from the surrounding diamond lattice that keep the damaged layers under compression are progressively reduced as the layers get closer to the sample surface. The electrical characterization of the channels strongly indicates that the ohmic conduction occurs in deep channels which are disconnected from the sample surface, with the exception of their endpoints, which are in electrical contact with surface metal electrodes. Moreover, the qualitative dependence of the channels conductivity as a function of the implantation fluence confirms that the critical damage density for graphitization in diamond with deep ion implantation is significantly higher than for shallow ion implantation.

**Acknowledgments**

The work of P. Olivero is supported by the "Accademia Nazionale dei Lincei – Compagnia di San Paolo" Nanotechnology grant, which is gratefully acknowledged. This work was supported by experiment "DANTE" of "Istituto Nazionale di Fisica Nucleare" (INFN).

**Tables**

Table 1

| Channel | Length (μm) | Width (μm) | Thickness (nm) | Resistance (MΩ) | Resistivity (Ω cm) |
|---|---|---|---|---|---|
| 2 | 922 | 16 | 145 | 3.5 | **0.9** |
| 3 | 712 | 28 | 195 | 1.5 | **1.1** |

Table 1: geometrical and electrical parameters taken from optical, profilometry and IV measurements to evaluate the resistivity of channels #2 and #3 after thermal annealing.

**Figure captions**

Fig. 1: TRIM Monte Carlo simulations of the damage density profile induced in diamond by 100 keV C at a fluence $F=3 \cdot 10^{15}$ cm$^{-2}$ [PRB_51_15711], 320 keV Xe at $F=3 \cdot 10^{15}$ cm$^{-2}$ [PRB_51_15711], 150 keV Ar at $F=1 \cdot 10^{15}$ cm$^{-2}$ [NIMB_32_145], 90 keV B at $F=3 \cdot 10^{15}$ cm$^{-2}$ [DRM_5_752], 2 MeV B at $F=3 \cdot 10^{15}$ cm$^{-2}$ [APL_71_1492], 2 MeV He at $F=4.6 \cdot 10^{15}$ cm$^{-2}$ [DRM_15_1714] and 6 MeV C at $F=1.5 \cdot 10^{16}$ cm$^{-2}$ (present work). The graphitization thresholds for shallow and deep implantations are also reported in dashed and dot-dashed lines, respectively.

Fig. 2: Schematics of the three-dimensional masking technique adopted to control the penetration depth of implanted ions. A semi-circular mask (in green) is positioned on the diamond surface, so that the depth at which the heavily damaged layer (in black) is formed is modulated at the end of the ion beam scan (Fig. 2a). After the removal of the mask, the damaged layer is connected with the sample surface at its endpoint (Fig. 2b).

Fig. 3: Optical images in reflection (a, b) and schematics (c) of the diamond sample after masking. Fig. 3a: the semi-spherical gold contacts are visible on their respective adhesion Cr-Au pads (A, B, C, D). Fig. 3b: zoom image of the area highlighted in red in Fig. 3a. Fig. 3c: schematics of the sample cross section, highlighting the adhesion Cr layer (in red), above which the Au adhesion contact and semi-spherical mask are deposited (in green); the image is not to scale.

Fig. 4: Optical images in transmission of the implanted sample before thermal annealing; Fig. 4b displays a zoom image of the area highlighted in red in Fig. 4a. The implanted strip areas are clearly visible in transmission due to the opaqueness of the damaged diamond layer. Three channels (1, 2 and 3 in Fig. 4a) were implanted at increasing fluence between the contact masks A-B, B-D and C-D, respectively. Test implantations are visible in the upper part of the sample.

Fig. 5: Interferometric profilometry map of the channel #2 after thermal annealing (a) and typical cross sectional profile of the same channel (b); the pronounced swelling localized at the implanted area is clearly visible. A 16 μm length bar is visible in (b), corresponding to the channel width highlighted in Fig. 4b.

Fig. 6: AFM map collected in tapping mode at the endpoint where a buried channel is connected with the sample surface (a); the arrows in Fig. 6a indicate the directions along which the profiles reported in b, c and d are collected. The increasing swelling thickness at the channel endpoint is clearly visible.

Fig. 7: IV characteristics of as-implanted channels, in linear-linear (a) and linear-logarithmic (b) scales. The different curves are relevant to different measure points (see Fig. 4a): the red curve corresponds to channel #3 (connecting points C and D), the blue line to channel #2 (B-D), and the green line to channel #1 (A-B). The orange line corresponds to test measurement at contacts E-F.

Fig. 8: IV characteristics of annealed channels, in linear-linear (a) and linear-logarithmic (b) scales. The different curves are relevant to different measure points (see Fig. 4a): the red curve corresponds to channel #3 (connecting points C and D), the blue line to channel #2 (B-D), and the green line to channel #1 (A-B). The orange line correspond to test measurement at contacts E-F.

Fig. 9: TRIM Monte Carlo simulations of the damage density profile induced in diamond by 6 MeV C at fluences of $6.3 \cdot 10^{15}$ cm$^{-2}$ (green line), $1.1 \cdot 10^{16}$ cm$^{-2}$ (blue line) and $1.7 \cdot 10^{16}$ cm$^{-2}$ (red line). The graphitization threshold is also reported (dashed line).

Fig. 10: Optical image in transmission of the implanted sample after annealing and removal of the mask contacts. Channel #1 appears more transparent than channel #2, indicating a partial recovery of the diamond structure during the thermal processing.

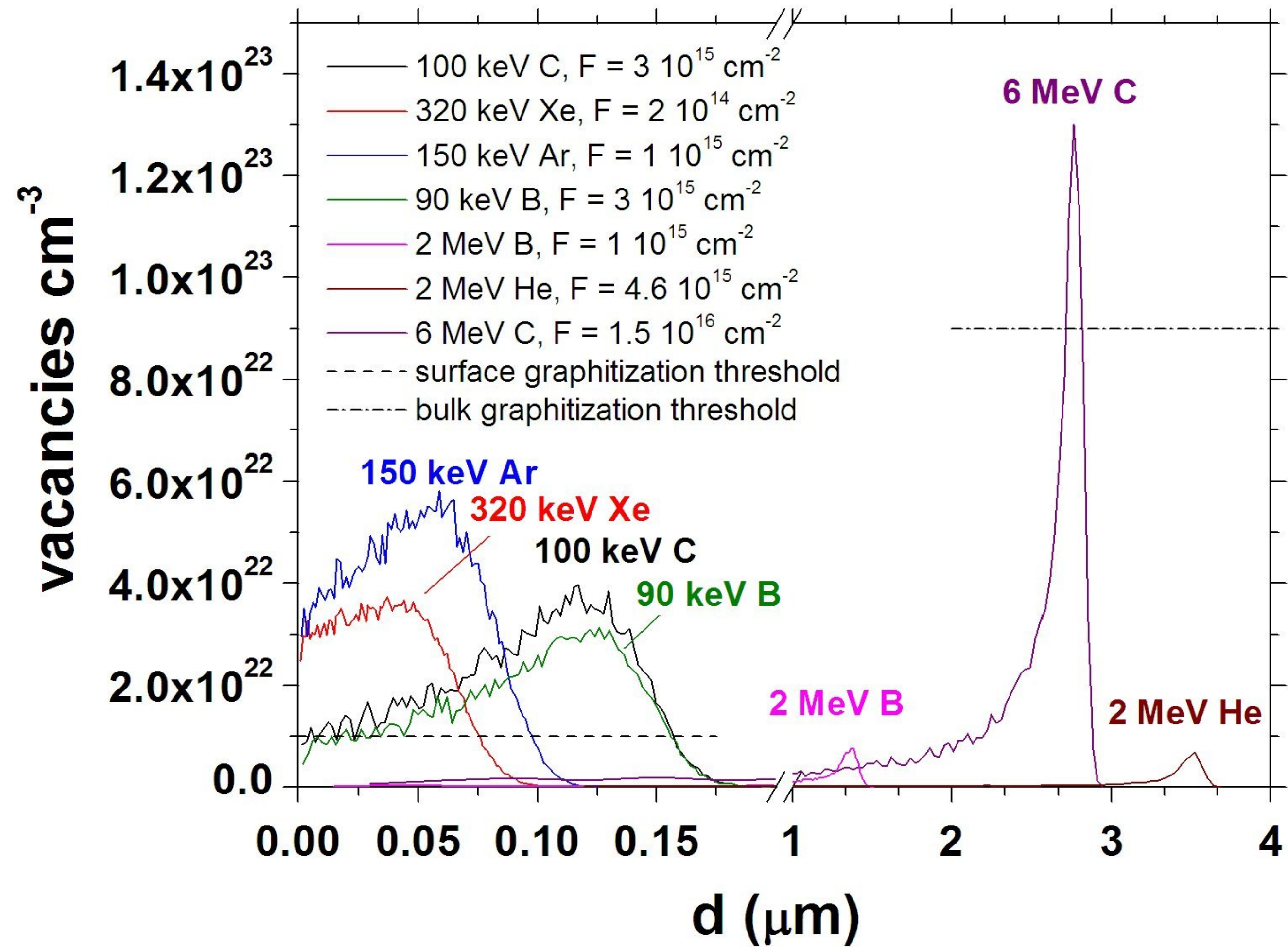

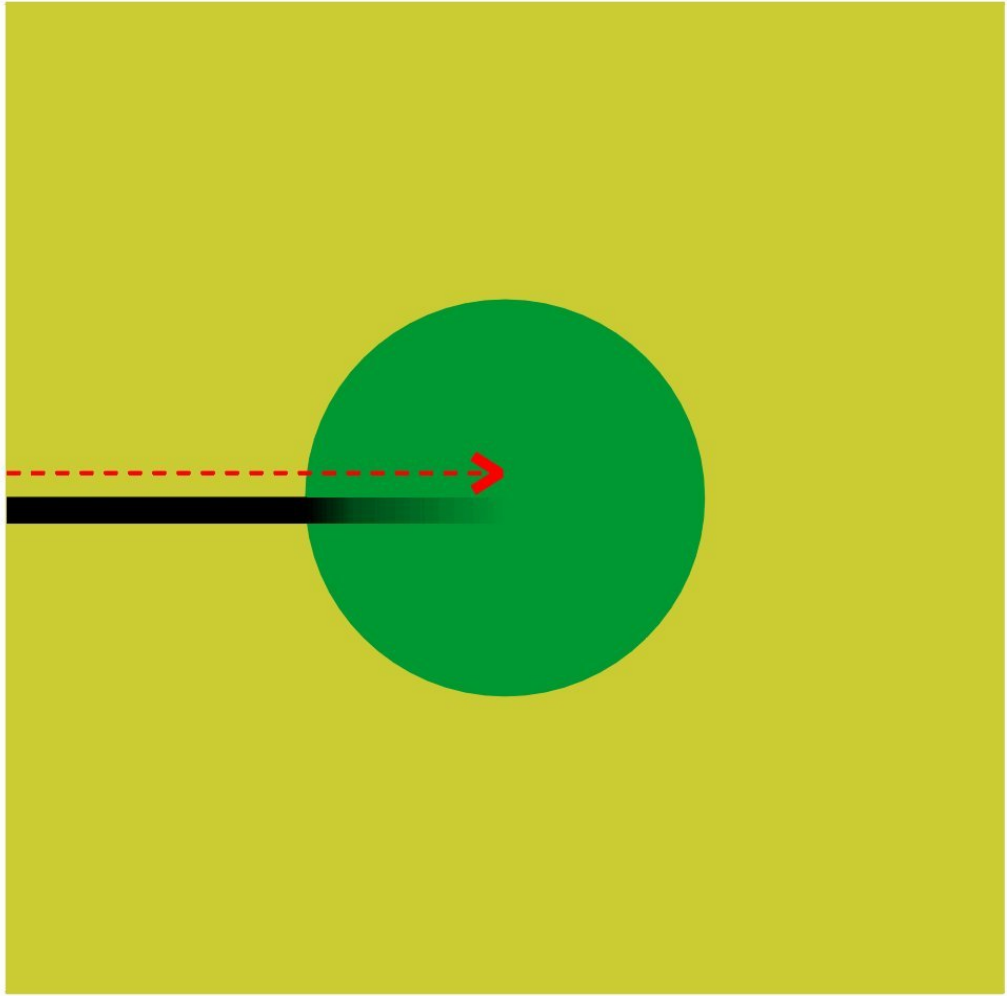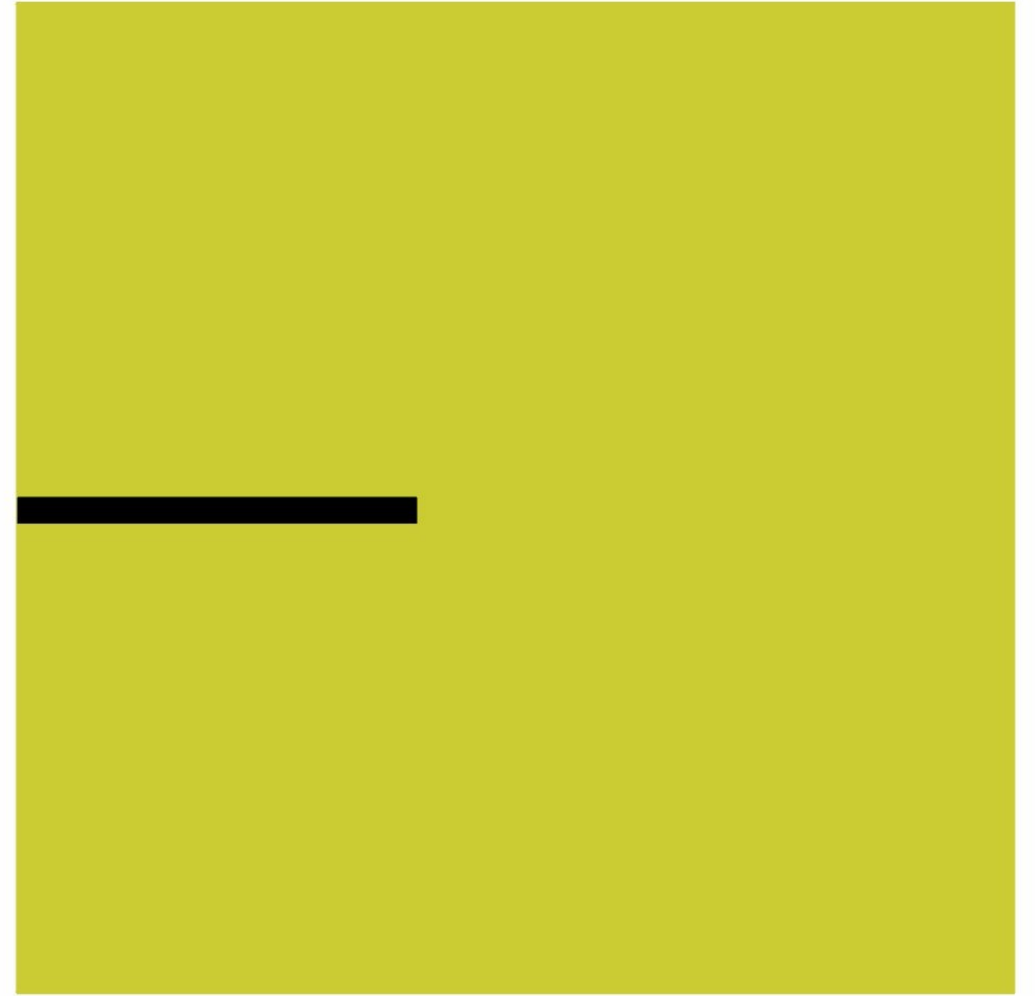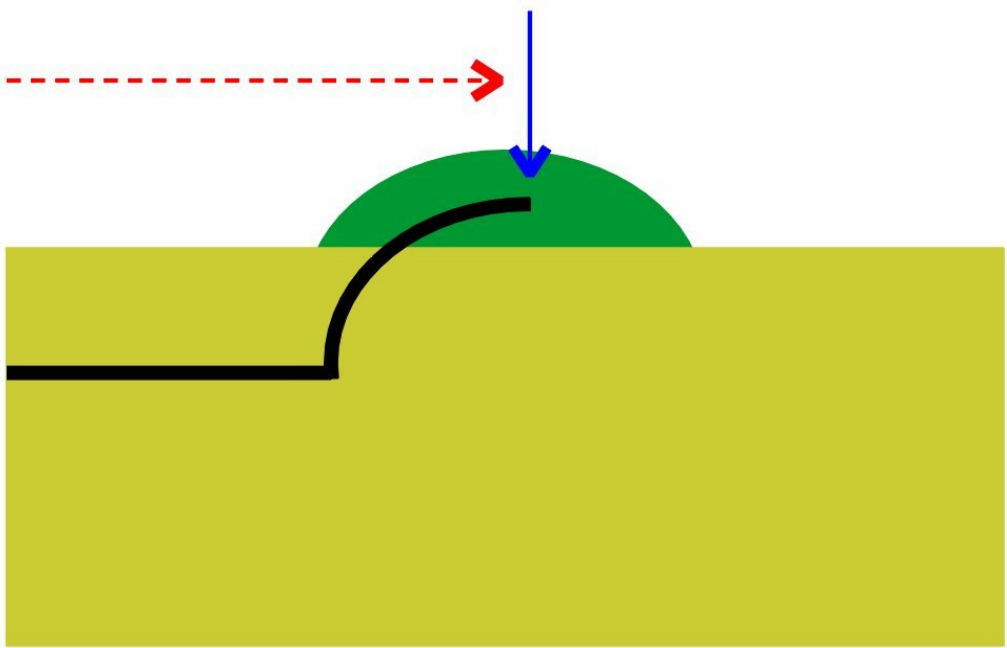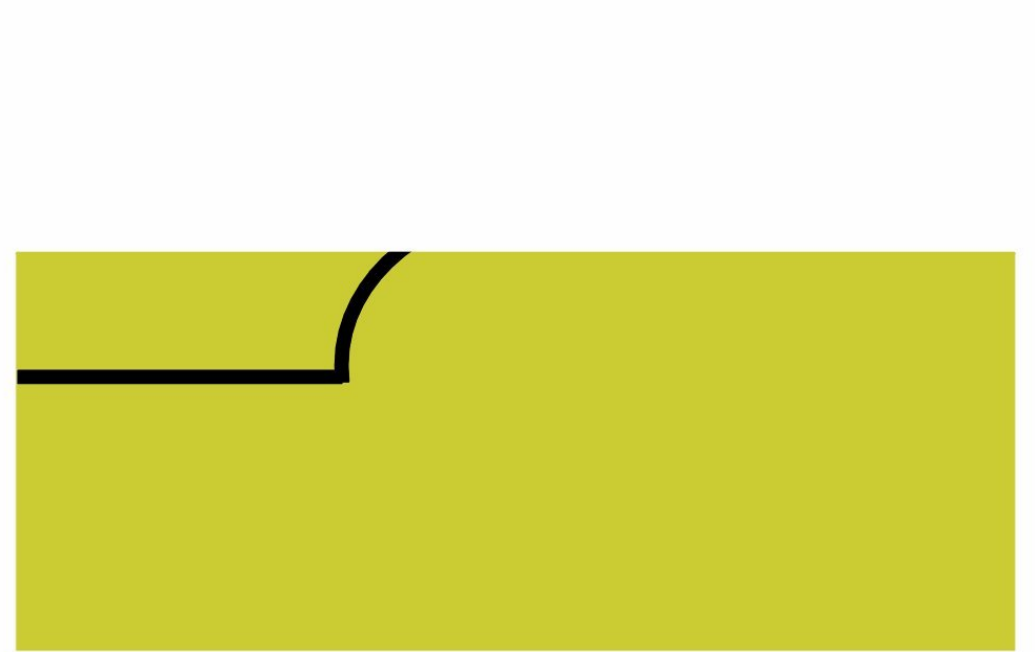

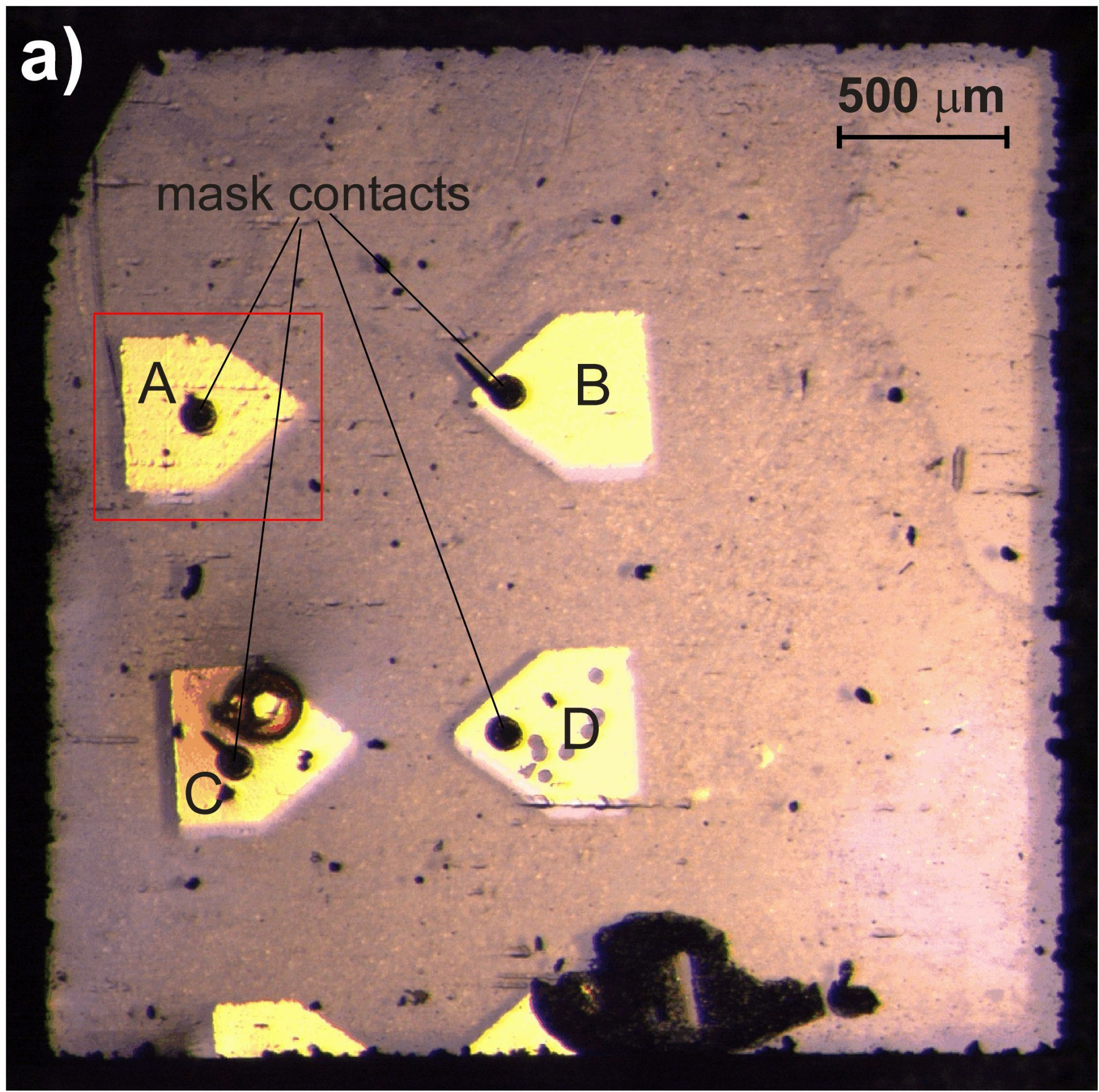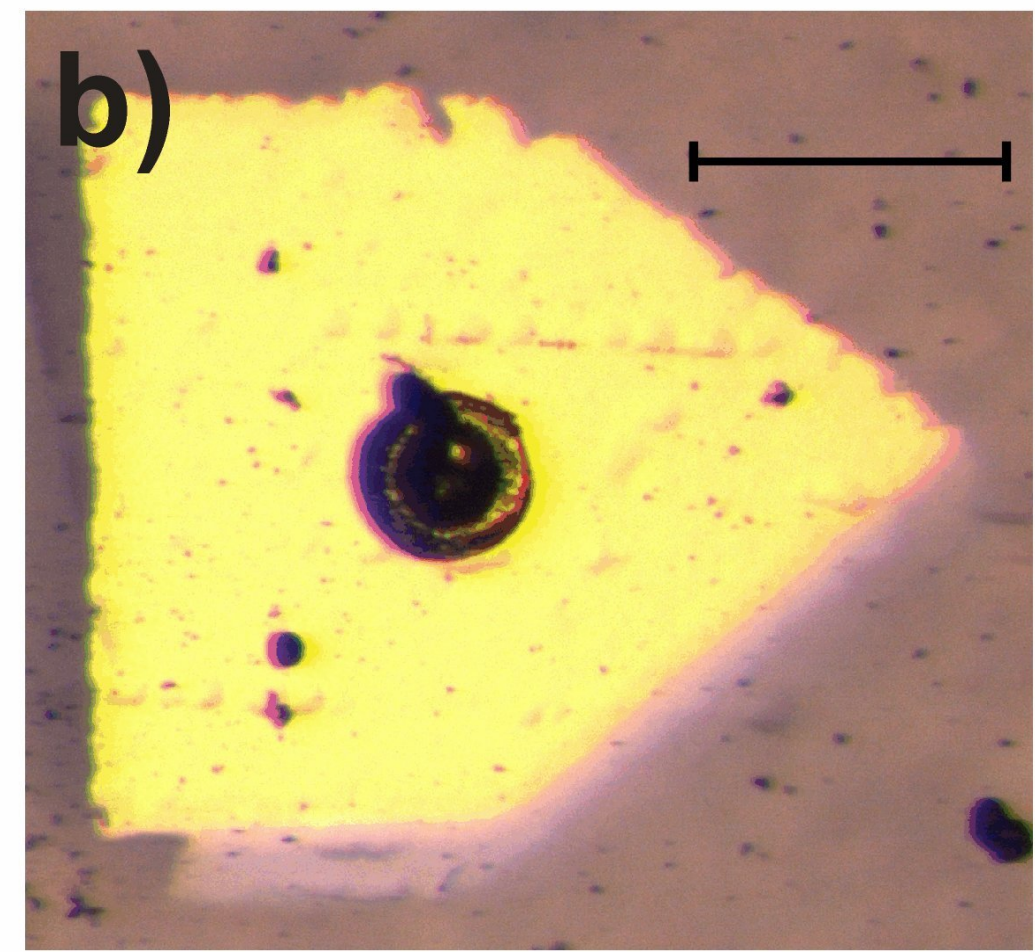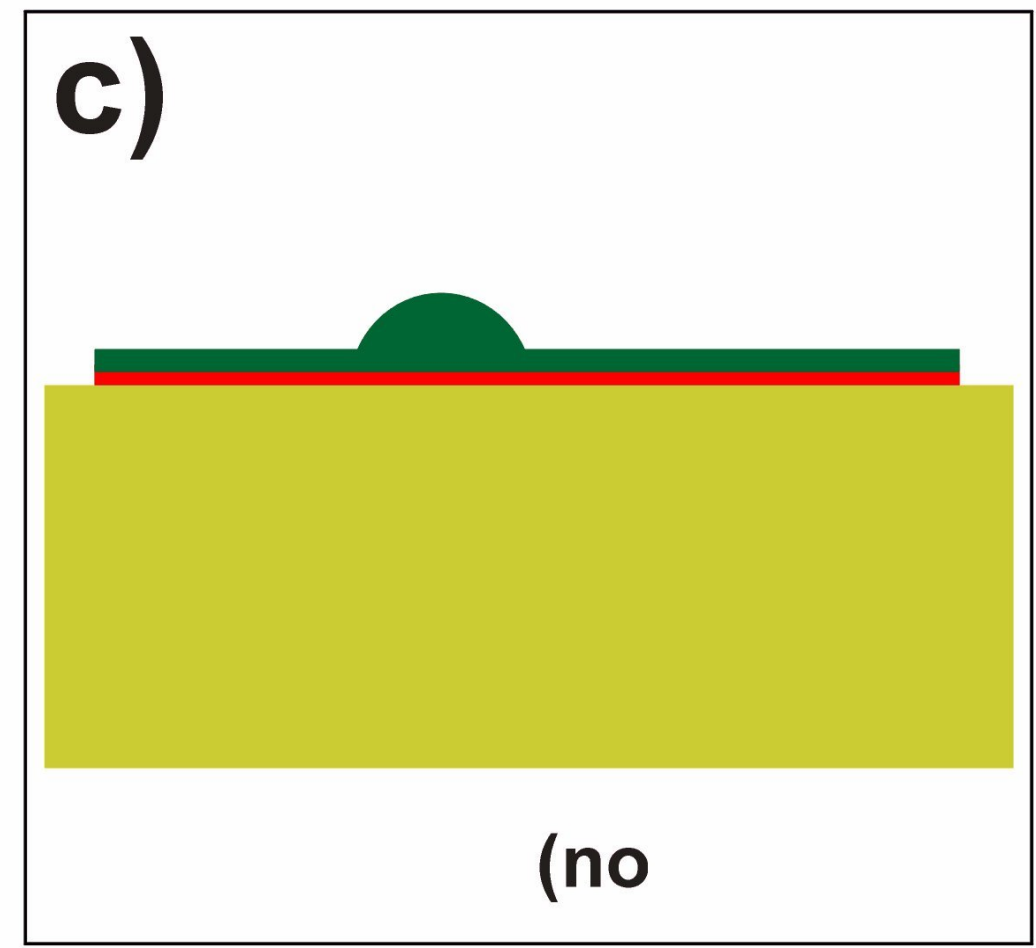

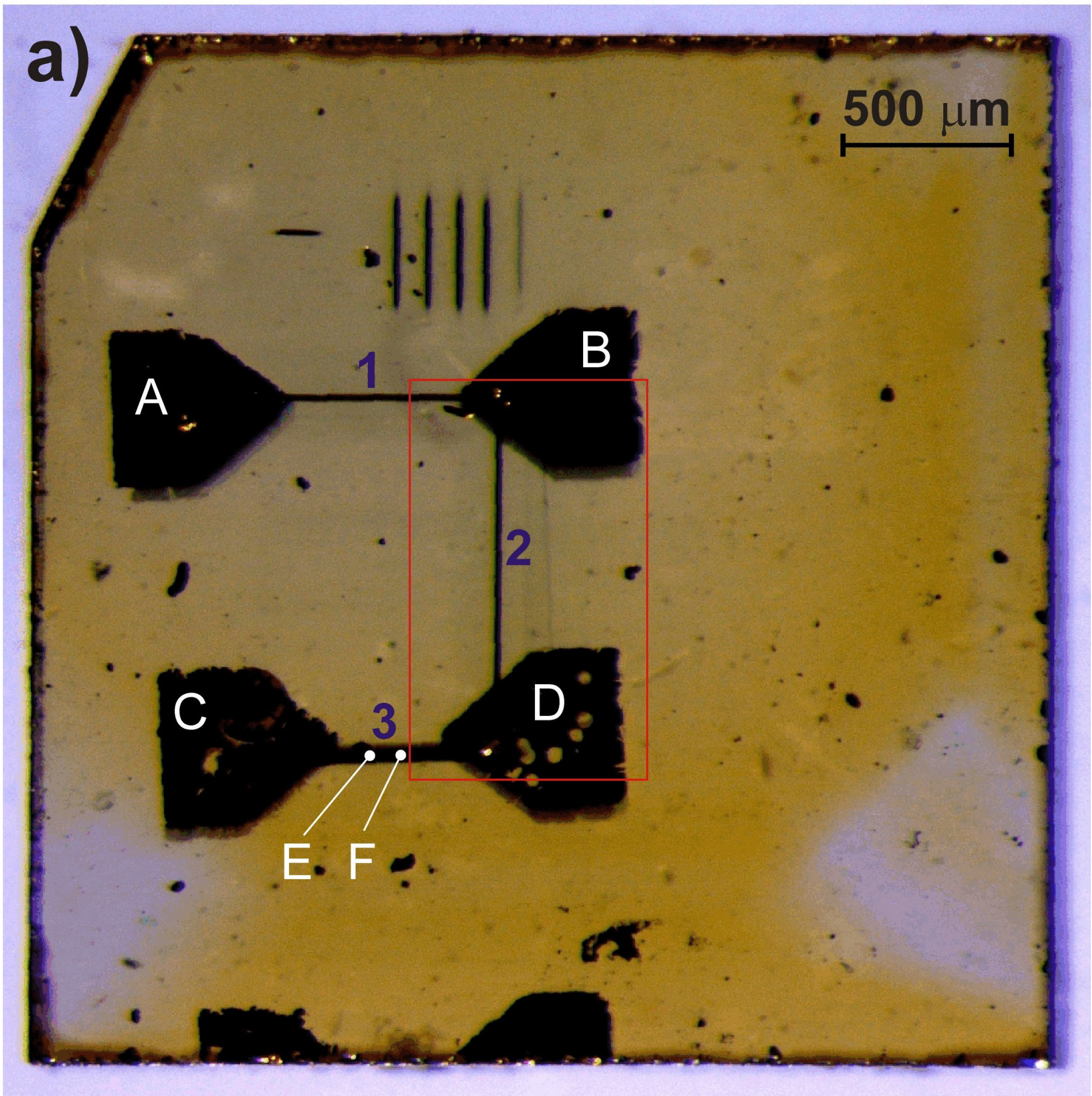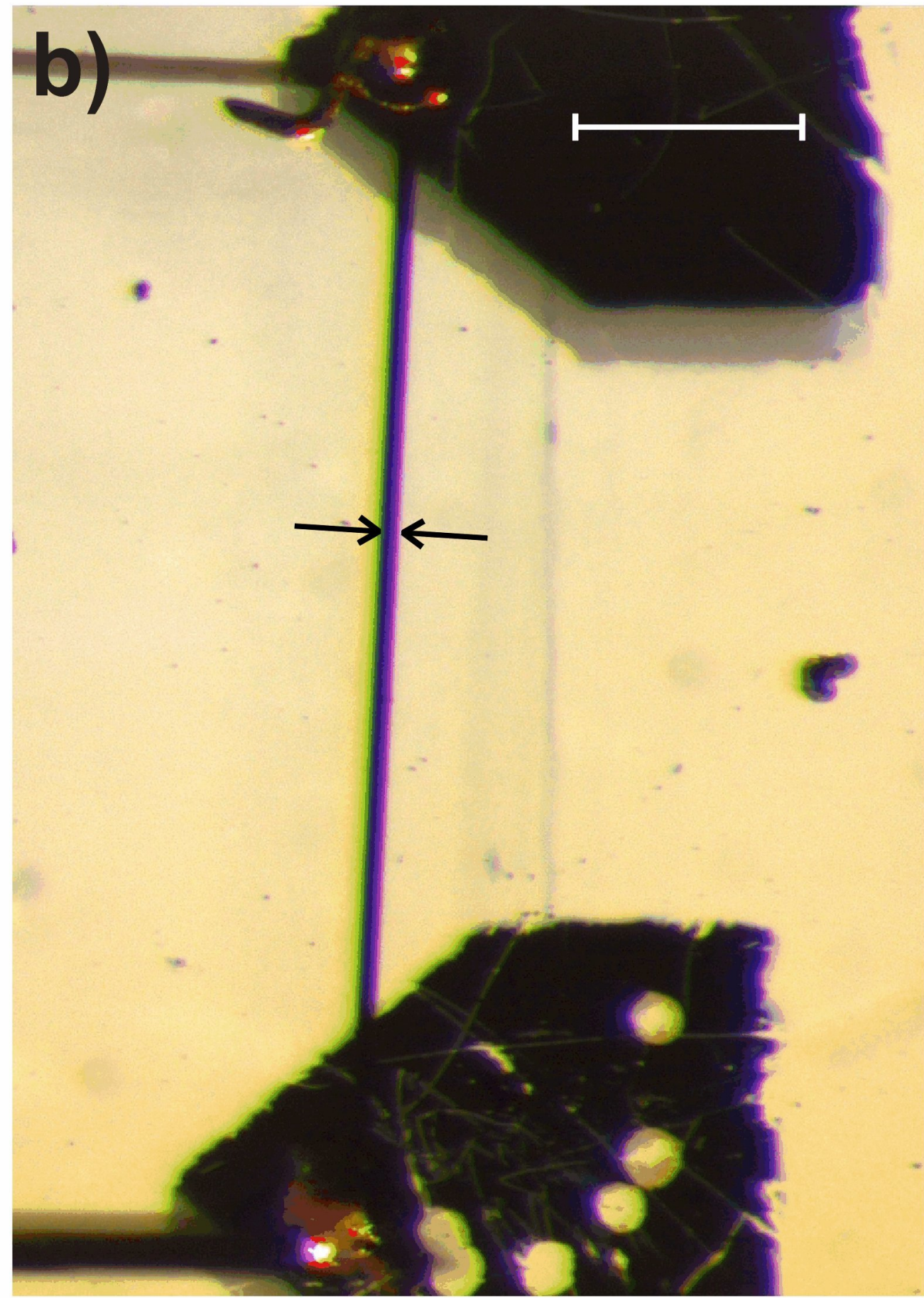

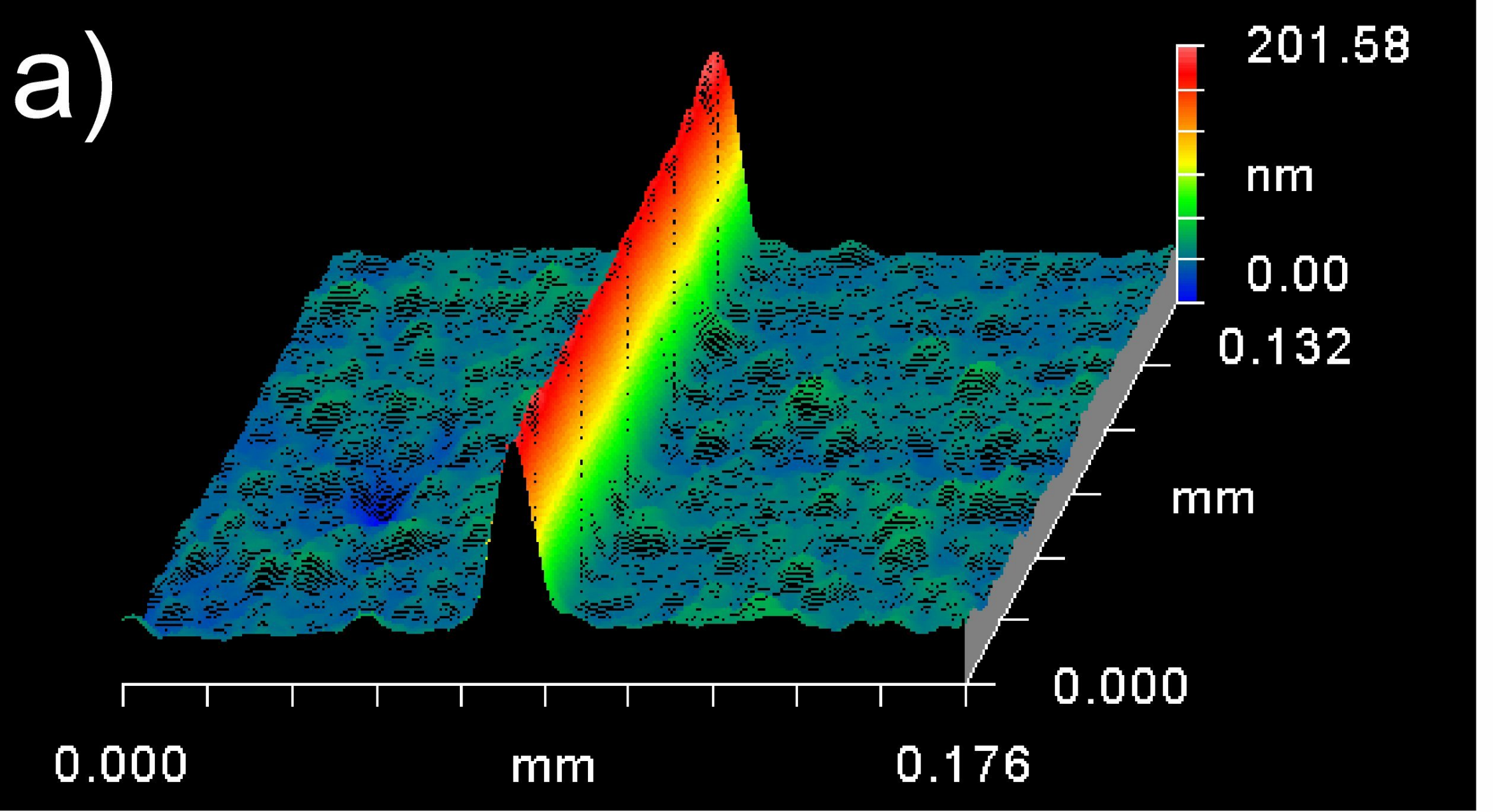
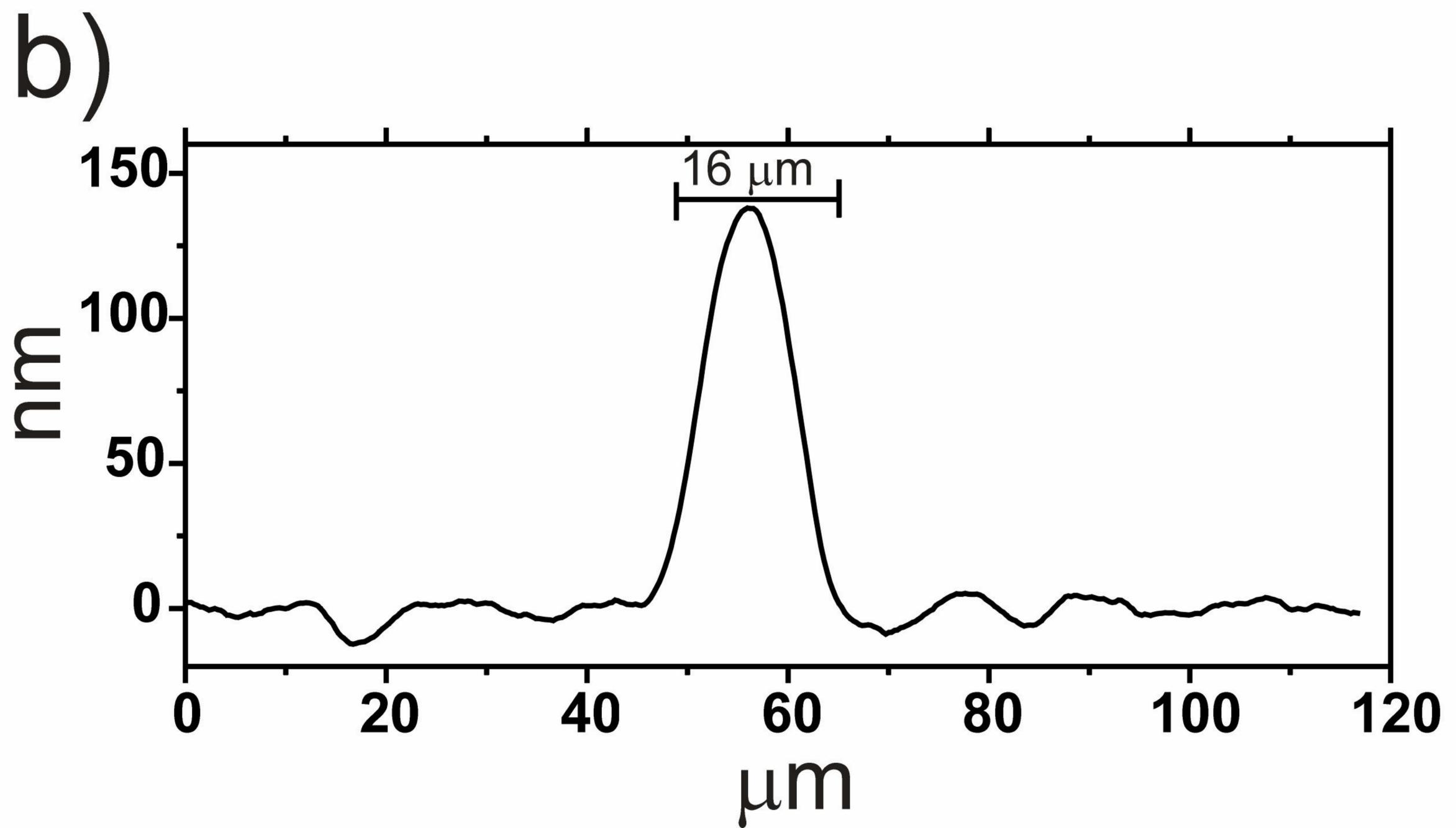

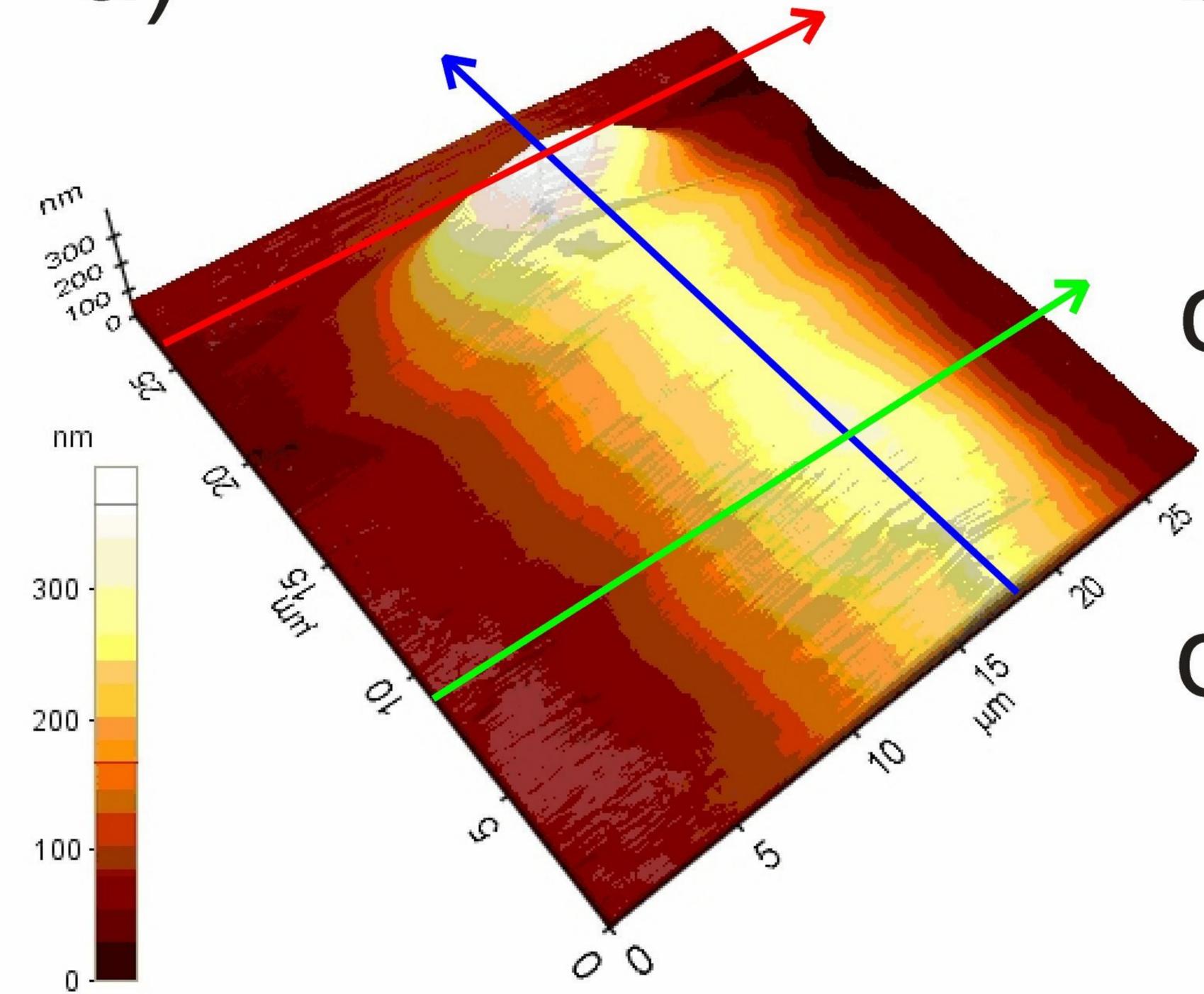
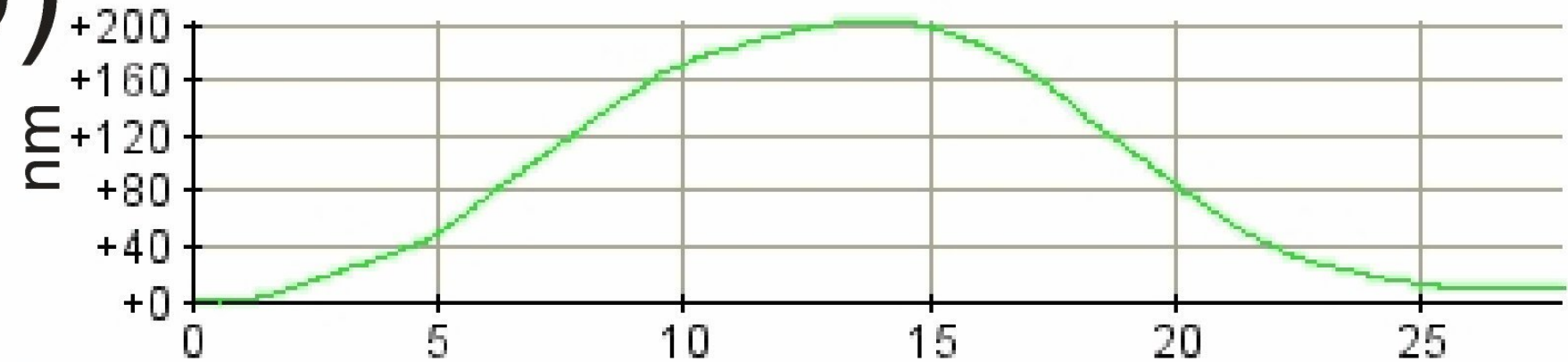
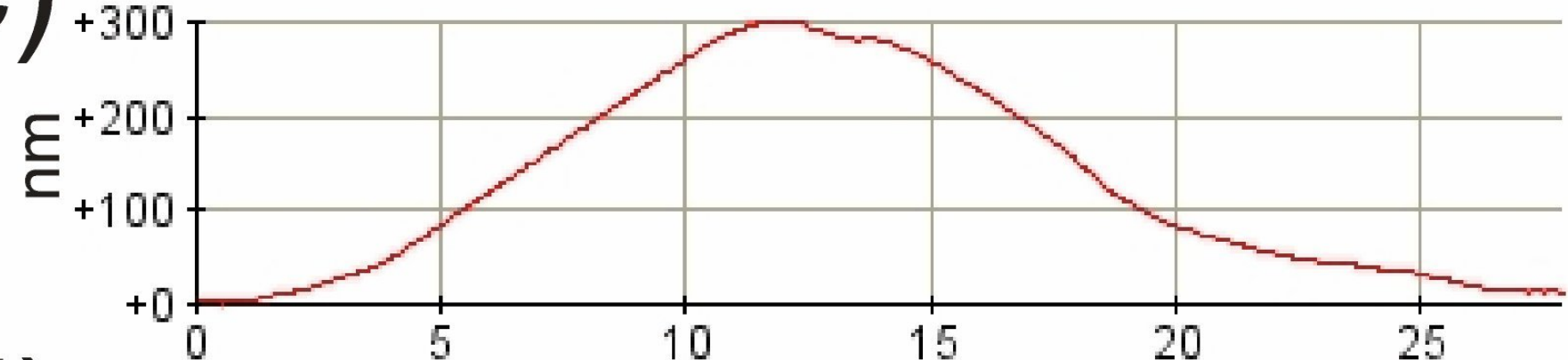
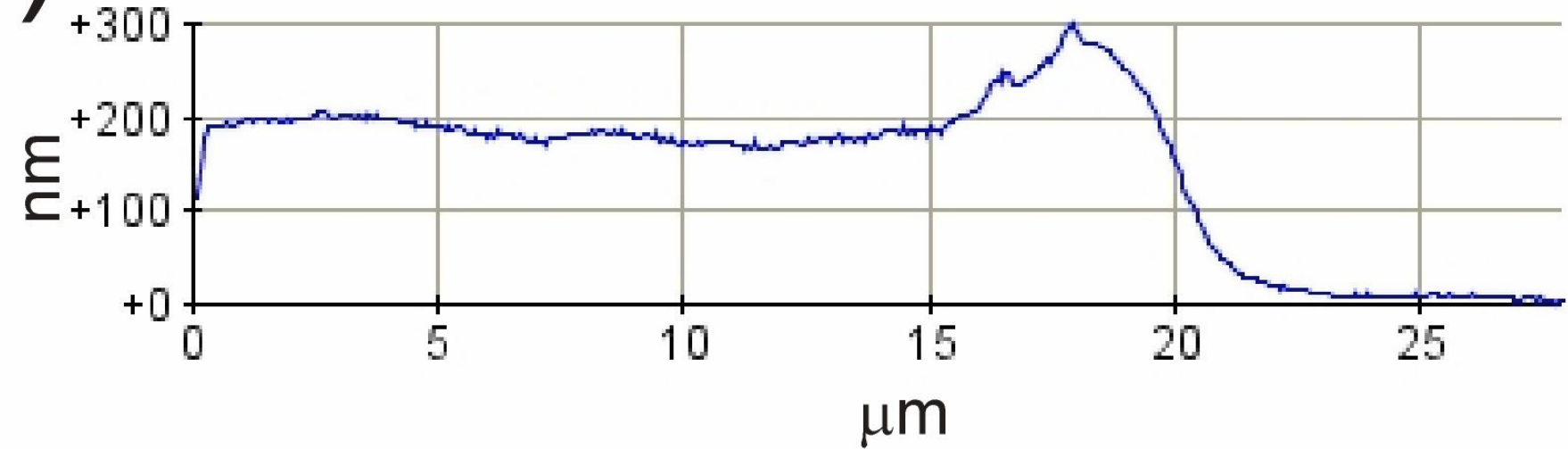

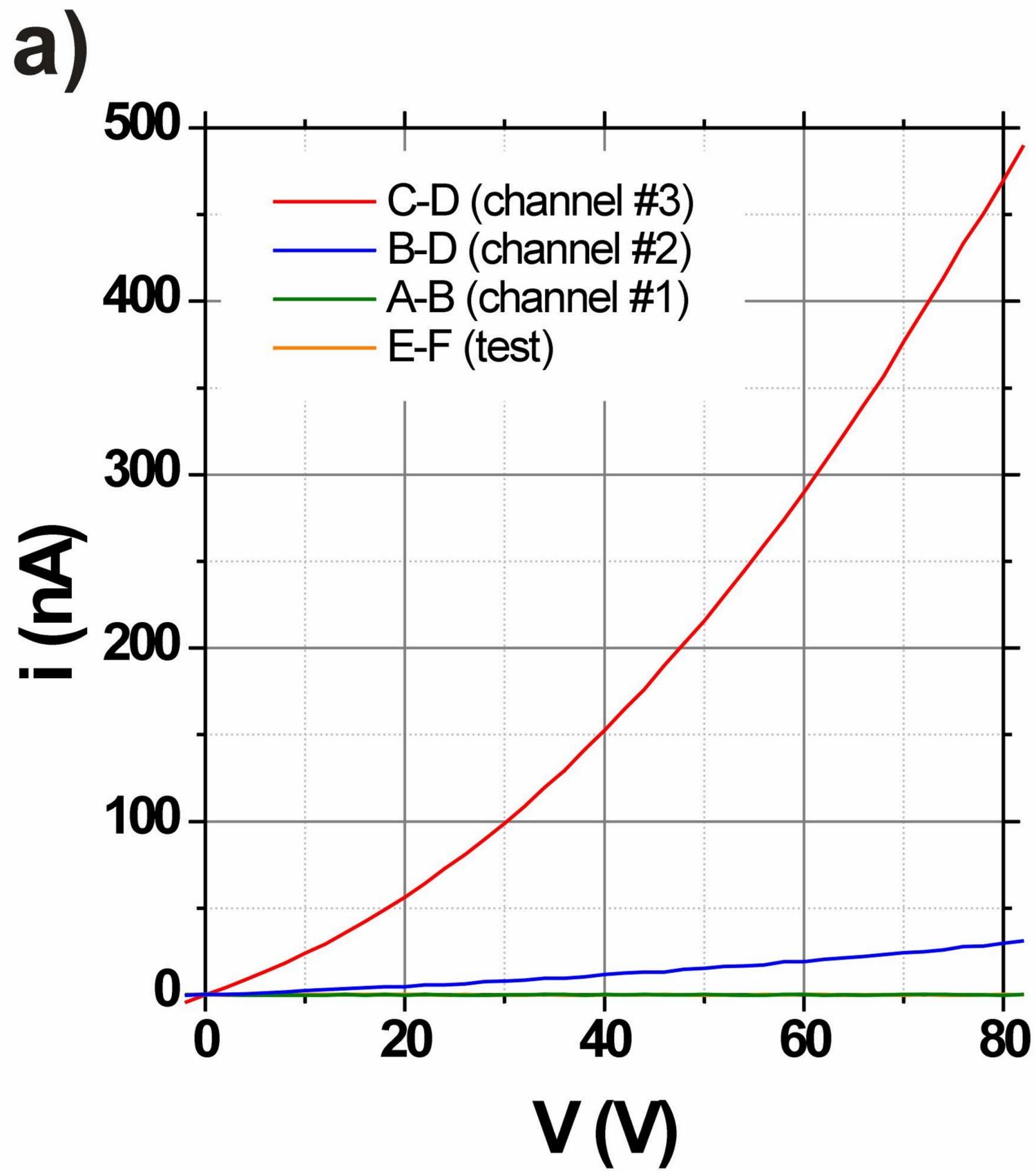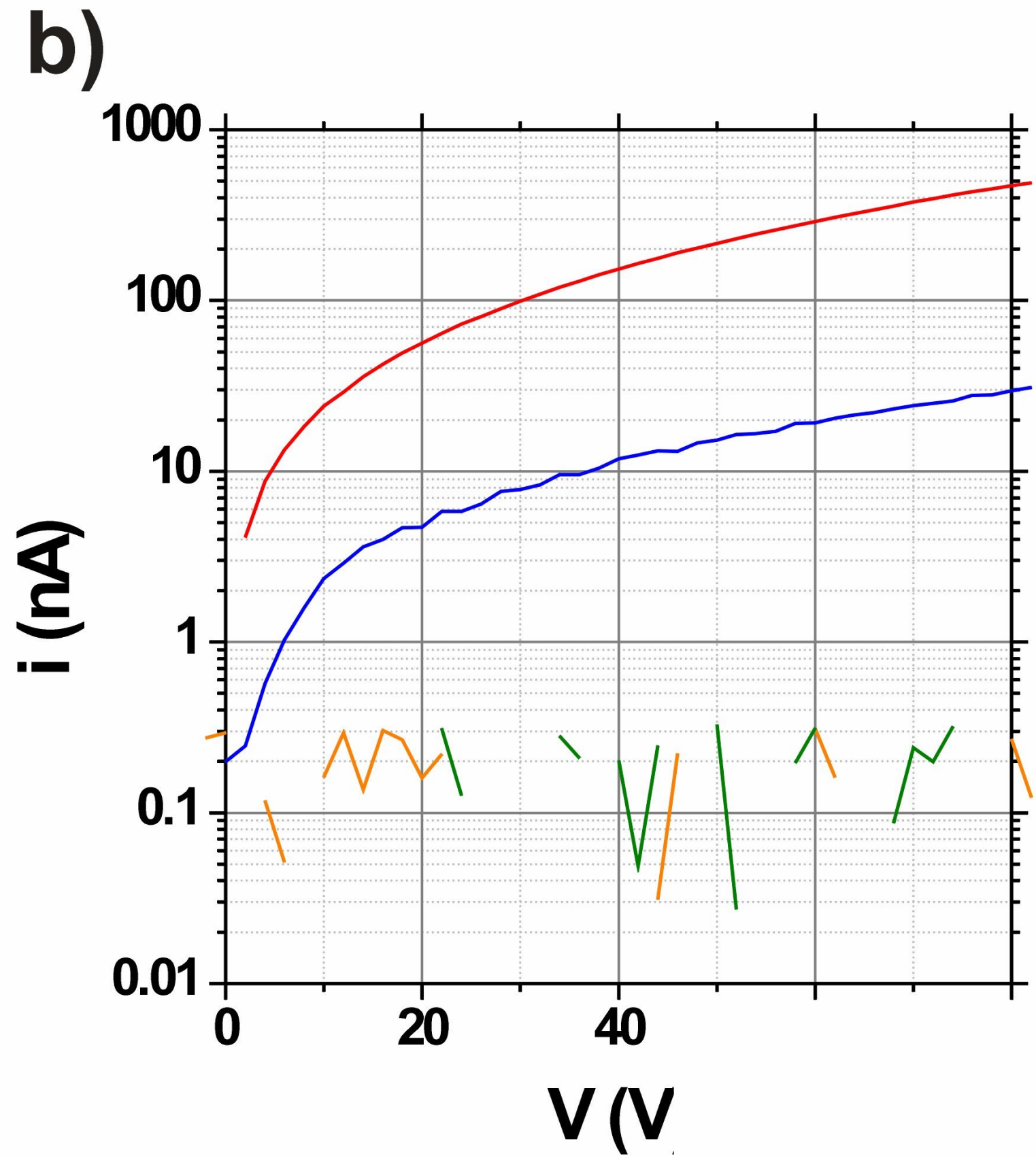

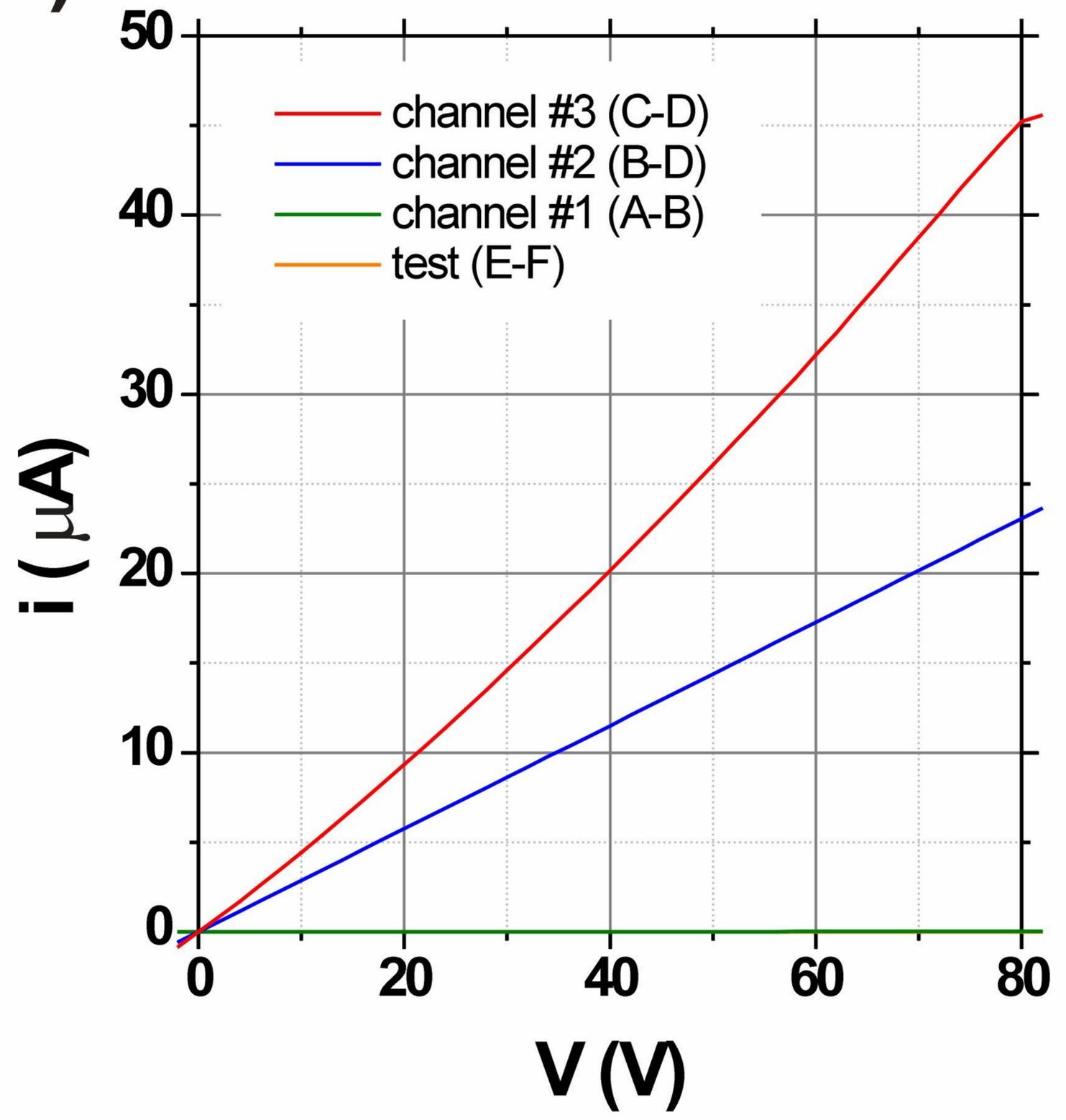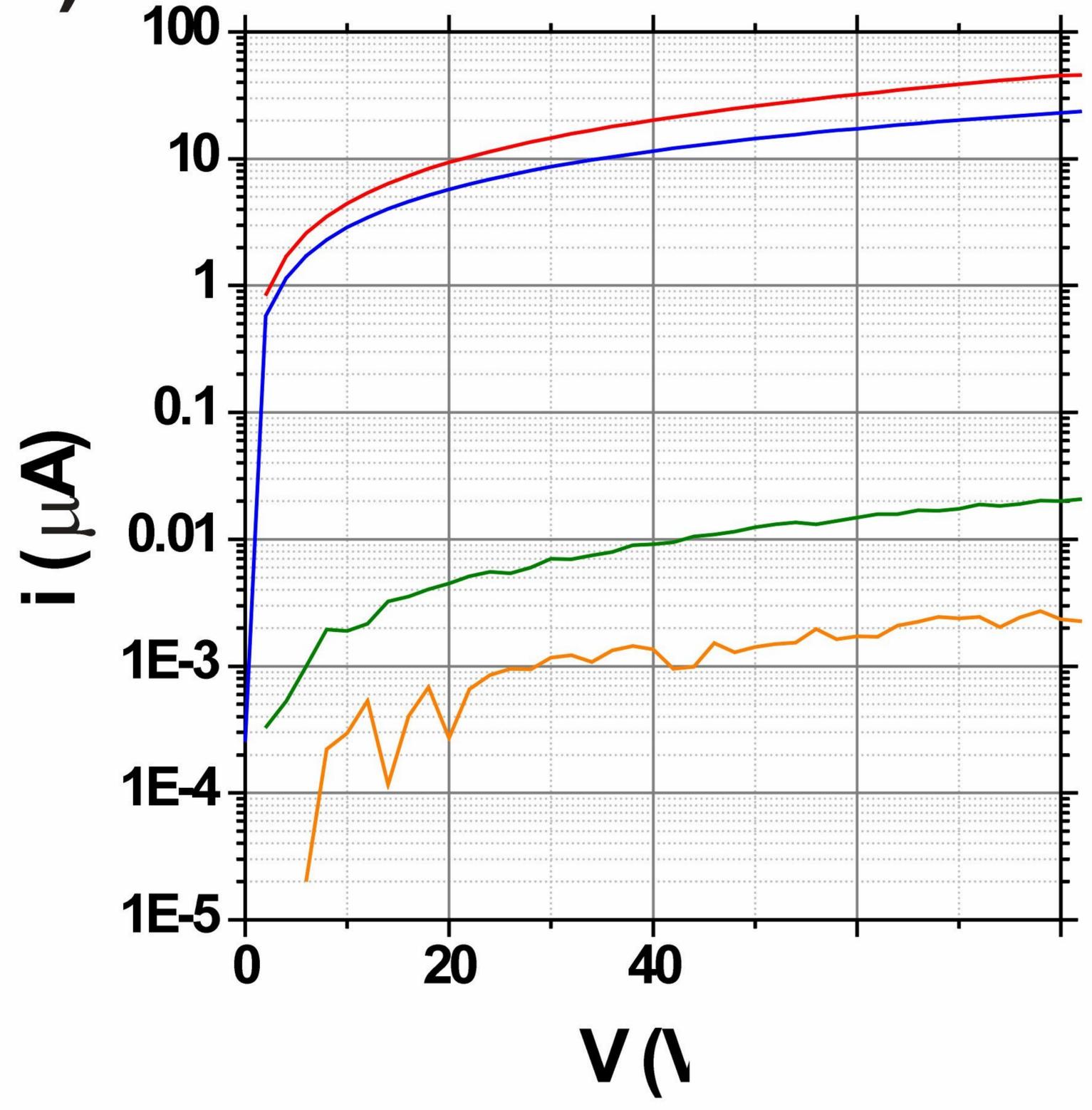

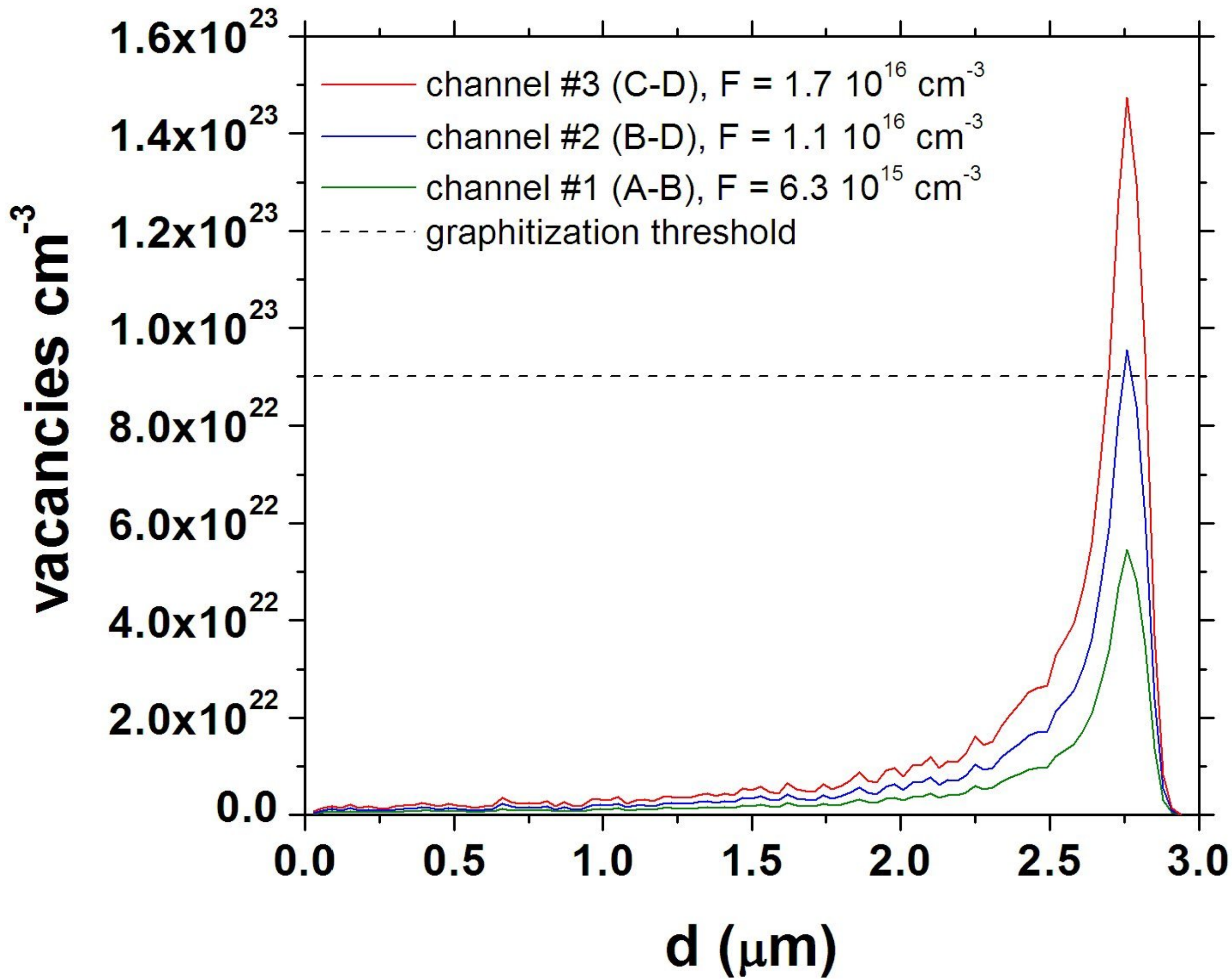

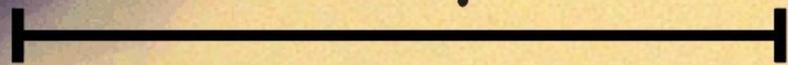